%% file: main.tex
\documentclass[sigconf,nonacm]{acmart}

\usepackage{enumitem}
\usepackage{fontawesome}
 \usepackage{array,multirow,graphicx}
 \usepackage{float}
\usepackage{todonotes}
\usepackage{hyphenat}

\usepackage{acronym} 

\usepackage{forest}

\AtBeginDocument{%
  }

\setcopyright{acmcopyright}
\copyrightyear{2023}
\acmYear{2023}
\acmDOI{XXXXXXX.XXXXXXX}

\acmConference[JCDL '23]{ACM/IEEE Joint Conference on Digital Libraries}{June 26 -- 30 2023}{Santa Fe, New Mexico, USA}

\acmPrice{15.00}
\acmISBN{978-1-4503-XXXX-X/23/06}

\begin{document}

\input{acronyms.tex}

\title{Bibliometric Data Fusion for Biomedical Information Retrieval}

\author{Timo Breuer}
\orcid{0000-0002-1765-2449}
\affiliation{
  \institution{TH Köln - University of Applied Sciences}
  \country{Germany}
}
\email{timo.breuer@th-koeln.de}

\author{Christin Katharina Kreutz}
\orcid{0000-0002-5075-7699}
\affiliation{
  \institution{TH Köln - University of Applied Sciences}
  \country{Germany}
}
\email{christin.kreutz@th-koeln.de}

\author{Philipp Schaer}
\orcid{0000-0002-8817-4632}
\affiliation{
  \institution{TH Köln - University of Applied Sciences}
  \country{Germany}
}
\email{philipp.schaer@th-koeln.de}

\author{Dirk Tunger}
\orcid{0000-0001-6383-9194}
\affiliation{%
  \institution{Forschungszentrum Jülich}
  \country{Germany}
}
\email{d.tunger@fz-juelich.de}

\renewcommand{\shortauthors}{Breuer, Kreutz, Schaer, and Tunger}

\begin{abstract}
Digital libraries in the scientific domain provide users access to a wide range of information to satisfy their diverse information needs. Here, ranking results play a crucial role in users' satisfaction. Exploiting bibliometric metadata, e.g., publications' citation counts or bibliometric indicators in general, for automatically identifying the most relevant results can boost retrieval performance. This work proposes bibliometric data fusion, which enriches existing systems' results by incorporating bibliometric metadata such as citations or altmetrics. Our results on three biomedical retrieval benchmarks from \ac{TREC-PM} show that bibliometric data fusion is a promising approach to improve retrieval performance in terms of \ac{nDCG} and \ac{AP}, at the cost of the \ac{P@10} rate. Patient users especially profit from this lightweight, data-sparse technique that applies to any digital library.
\end{abstract}



\maketitle

\section{Introduction}
\label{sec:intro} 

Metadata curation of digital libraries helps to improve the retrieval performance \cite{DBLP:conf/jcdl/GhosalCSE0B19} and is beneficial for searching the body of literature. However, manual metadata annotations by domain experts are costly and do not scale well. As a special type, bibliometric metadata does not require any explicit labeling by domain experts as it results from the meta-analysis of scholarly communication and is implicitly based on the reception by the scientific community.

It was shown that bibliometric metadata correlates with manual relevance labels, as known in test collections from the domain of biomedicine or physics \cite{DBLP:journals/scientometrics/BreuerST22}. Furthermore, there is a high correlation between bibliometrics and documents with positive relevance judgments. While it is controversially discussed to which extent bibliometric data reflects topical relevance, this kind of metadata can be considered an implicit relevance signal that can potentially improve retrieval performance.  However, previous research has shown that including bibliometric measures like citation rates in the retrieval process is not trivial \cite{DBLP:journals/ipm/Pao93}.

One suggestion on how to align different kinds of relevance signals or representations is the principle of polyrepresentation, introduced by Ingwersen \cite{ingwersen_cognitive_1996}. It is based on the idea that different retrieval models can be regarded as different perspectives on information retrieval. The principle suggests that different models retrieve different sets of information from the same collection and that these sets might include different representations of the same document. According to the principle of polyrepresentation, there is an increasing chance of relevant documents being retrieved, and combining these different representations improves retrieval performance compared to using only single representations alone.
Data fusion methods are well-known techniques in the field of meta-search, where multiple ranking outputs are combined for the sake of retrieval effectiveness \cite{DBLP:conf/sigir/AslamM01}. The principle of polyrepresentation and data fusion align on a conceptional level, and they can also be combined on the level of concrete retrieval systems \cite{DBLP:journals/jasis/LarsenIL09}. 

In this work, we combine bibliometric data with fusion methods and the pre-computed result lists (run files) submitted to the \ac{TREC-PM} Abstract task from 2017 to 2019 \cite{DBLP:conf/trec/RobertsDVHBLP17,DBLP:conf/trec/RobertsDVHBL18,DBLP:conf/trec/RobertsDVHBLPM19} to investigate the effect of including bibliometric indicators into the ranking of biomedical retrieval systems. Our experiments show that bibliometric information like citations or altmetrics has some discriminating power that can be beneficial for retrieval performance. However, bibliometric metadata alone does not include topical relevance criteria, which limits effective rankings. By  combining bibliometric information with topical relevance criteria, as they were implemented into the systems of \ac{TREC-PM}, we demonstrate how the retrieval performance of biomedical retrieval systems can be improved with data fusion techniques. Finally, we address the expected benefit for users of digital libraries when implementing bibliometric data fusion approaches into the search process. More precisely, we address the following research questions:

\begin{enumerate}[label=\textbf{RQ\arabic*},leftmargin=10mm]
  \item \emph{To what extent can bibliometric relevance signals be used as ranking criteria for biomedical information retrieval?} 
  \item \emph{Can bibliometric-enhanced data fusion methods improve the overall retrieval performance?} 
\end{enumerate}

The remainder includes the related work in Section \ref{sec:related_work}, covering polyrepresentation, biomedical information retrieval, and bibliometric measures in information retrieval. In Section \ref{sec:data_fusion}, we recapture fundamentals of data fusion and present our selected data fusion approaches. Section \ref{sec:dataset} introduces our dataset, which is followed by an outline of the methodology in Section \ref{sec:methodology}. Afterward, we present the corresponding experimental results that give answers to the research questions in Section \ref{sec:experimental_results}. Finally, we conclude in Section \ref{sec:conclusion}.

\newpage
\section{Related Work}
\label{sec:related_work}

Adjacent areas to this work are 
\textit{polyrepresentation in information retrieval}, \textit{biomedical information retrieval} in general, and using \textit{scientometric measures in information retrieval}.

\subsection{Polyrepresentation in Information Retrieval}

Polyrepresentation is a concept that has been developed as a result of a cognitive approach to information retrieval~\cite{ingwersen_cognitive_1996}.  In this framework, retrieval models represent the retrieval system developer's ideas and perspectives on information retrieval. From this point of view, each retrieval model is cognitively different from other retrieval models because it represents a unique conceptual and algorithmic interpretation of information retrieval.

According to the principle of polyrepresentation, different retrieval models retrieve different sets of information from the same collection when given the same retrieval task. However, some overlap occurs between different models. The nature of this overlap depends on the conceptual and algorithmic interpretation of similarity. It is important to note that a relatively high overlap of documents retrieved by different models does not automatically imply a similarity between the models. In fact, from a polyrepresentation perspective, a high overlap of documents may be an advantage. For example, suppose the fused retrieval models are dissimilar, meaning they interpret the original collection from quite different perspectives. In that case, the overlap signifies high odds of relevant documents being retrieved \cite{DBLP:journals/jasis/LarsenIL09}. According to Ingwersen \cite{ingwersen_cognitive_1996} the principle of polyrepresentation operates with two types of similarity/dissimilarity: ``cognitive dissimilarity'' when fundamentally different retrieval models are in action and ``functional difference'' when the fused entities are based on different versions of the same fundamental retrieval model.

An early work that used different retrieval models in combination for improved precision was presented by Croft and Thompson \cite{croft_imbox3r_1987}, which fused probabilistic and vector space models. 
Based on these studies, Larsen et al. \cite{DBLP:journals/jasis/LarsenIL09} reported on data fusion experiments using the four best-performing retrieval models from TREC 5, where three models were conceptually/algorithmically very different from one another and one was similar to one of the former. They concluded that the performance of data fusion on all possible combinations seems to depend on three factors: ``(a) the degree of conceptual/algorithmic dissimilarity between the constituent IR models, and (b) how equal and (c) well the component models perform.'' 
Other work based on the principle of  polyrepresentation focused on fusing different metadata representations, e.g., for query expansion \cite{schaer_extending_2012}, or digital library curation tasks like prioritizing different conferences for indexing \cite{10.1145/3197026.3197069}.

Ingwersen \cite{ingwersen_citations_2012} pointed out that citations and bibliographic references in scientific documents can be useful for document retrieval. These can be seen as ``footprints of information interaction'' that are crucial for scientific communication. Therefore, he argues that these should be exploited for document retrieval. 
Skov et al. \cite{skov_inter_2008} showed the general feasibility and positive influence on retrieval performance in a study on the Cystic Fibrosis test collection.
Belter \cite{belter_relevance_2017} proposes a method of ranking the relevance of citation-based search results based on seed documents and using citation relationship analysis for document ranking.

\subsection{Biomedical Information Retrieval}

There have been diverse efforts in the area of information retrieval systems focused on biomedical information needs:
As of April 2023, PubMed 
contains information on more than 35M publications from the MEDLINE collection, covering the biomedical domain. Information can be accessed via keyword search on specific fields such as titles, abstracts, MeSH\footnote{Medical Subject Headings, a taxonomy of biomedical concepts.} terms, or author names. 
Madaan~\cite{DBLP:journals/pvldb/Madaan13} proposed a domain-specific query language that was supposed to be used by current patients, thus novice users, as well as medical staff, thus domain-experts. The system was intended to retrieve and summarize the literature on cases related to current ones. 
Afsar et al.~\cite{DBLP:conf/flairs/AfsarCF21} present a paper recommendation system directed at patients to support their decision-making in medical treatments. 
Related to these works, the so-called narrative information access~\cite{DBLP:conf/jcdl/KrollPPB22}, part of the pharmaceutical digital library PubPharm, can also be used to answer queries related to diseases, treatments, and genes. Here, queries need to be formulated as triples.

From 2014 onwards, TREC focused on the previously under-explored research area of medical information retrieval by introducing the Clinical Decision track \cite{DBLP:conf/trec/SimpsonVH14}. From 2017, these efforts were continued as Precision Medicine~\cite{DBLP:conf/trec/RobertsDVHBLP17,DBLP:conf/trec/RobertsDVHBL18,DBLP:conf/trec/RobertsDVHBLPM19}. Recently, the focus is clinical trials since 2021~\cite{TREC22}.

\subsection{Bibliometrics in Information Retrieval}

Garfield~\cite{Garfield1964} developed a model of a science index, which made it possible for the first time not only to search for literature bibliographically or thematically, but also to find relevant publications through citation analyses. This marked the birth of bibliometrics and the Science Citation Index, which is still used today as part of the Web of Science. Over the years, citations became a currency in many scientific areas.

Opposed to \textit{explicit} \textbf{editorial relevance judgments}, bibliometric measures are a more \textit{implicit} type of relevance signals~\cite{DBLP:journals/scientometrics/BreuerST22}. As outlined by Voorhees \cite{voorhees_trec_2007}, the annotation process of topical relevance is guided by some text-based descriptions of the information need, which can usually be found in the topic files of a test collection. In this way, the decision behind the relevance label becomes more transparent and can often be determined by concrete criteria. 

Similarly, citing a publication also signals its overall relevance or quality, but in comparison, it is less transparent and explicit than an editorial label. For instance, a higher \textbf{number of citations} does not imply topical relevance by all means~\cite{ingwersen_bibliometrics/scientometrics_2012}. Fisher and Naumer~\cite{fisher_information_2006} point out criteria that could lead to a citation, including trustworthiness, contact, access or convenience, inexpensiveness, and ease of use, among others. Citations and particularly citation networks are helpful for cross-language recommendations of publications \cite{DBLP:conf/jcdl/JiangLL18}. However, citations can also be influenced by biases caused, for instance, by the affiliation of the authors \cite{DBLP:conf/jcdl/Nishioka0S22}. Moreover, not all citations are equally important. Some references have a larger impact on a study, while others only fall into the scope of the broader context. Hassan et al. \cite{DBLP:conf/jcdl/HassanAH17} propose machine learning classifiers to distinguish between important and less important citations in a paper. Citations provide valuable context information. As shown by Kehoe and Torvik~\cite{DBLP:conf/jcdl/KehoeT16}, citations can also help to estimate MeSH terms. Likewise, citations can be exploited for journal recommendations when combined with full-text information~\cite{DBLP:conf/jcdl/GhosalCSE0B19}.

\textbf{Altmetrics} complement traditional bibliometrics with citation statistics from social media and other online media \cite{DBLP:conf/jcdl/ShakeelAKSL21}. Thus, altmetrics can be compared to the introduction of the Science Citation Index, which enabled scientists to track where they have been cited for the first time. The only difference is that these ``citations'' are called news items, blog posts, likes, reads, shares, or readerships. Altmetrics make scientific impact visible more quickly than traditional bibliometrics because they evolve more quickly and dynamically. As shown by Shakeel et al.~\cite{DBLP:conf/jcdl/ShakeelAKLS22}, altmetrics correlate with citations, complement them, and can compensate for the citation bias.

Nishioka and Färber \cite{DBLP:conf/jcdl/Nishioka020} analyzed how open access types impact citations and altmetrics. They found that open-access articles receive higher citations than closed or gold articles. Breuer et al. \cite{DBLP:journals/scientometrics/BreuerST22} examined the connection between relevance assessments, citations, and altmetrics. It was found that the connecting element of these three dimensions is relevance. The corresponding dataset was compiled as a reusable artifact, covering all of the previously described (boldfaced) bibliometric measures, and is described in Section \ref{sec:methodology}. 

Scientific publications aim to contribute to state of the art in a particular field. For this purpose, reference is made to previous papers in this field, and these papers are cited respectively, indicating the relevance of these papers and a thriving flow of knowledge. For the original papers, citations are generated this way, cumulated at the paper level. At the journal level, the number of citations a publication achieved on average can be quantified, which is known as the \textbf{impact (factor)}. Keselman \cite{DBLP:conf/jcdl/Keselman19} treated venue authorship as a regression problem and proposed a method that can be used to evaluate the quality of venues. At the author assessment level, the evaluations showed that the venue-based method yielded comparable quality estimates to citation-based indicators. 

Similar to the impact factor, there is another measures on the journal level. The \textbf{research level} describes the journal's research orientation on four levels: clinical observation (applied technology, level 1), clinical mix (engineering-technological mix, level 2), clinical investigation (applied research, level 3) and basic research (level 4)~\cite{Narin1976,Boyack2014}. Boyack et al.~\cite{Boyack2014} classify single papers to these levels, e.g., by using titles and abstracts.

\section{Data Fusion}
\label{sec:data_fusion}

\begin{figure}[!t]
    \centering
    \resizebox{0.4\textwidth}{!}{
    \begin{forest}
      for tree={
      },
      [Data Fusion, for children
        [Rank-based
          [\textbf{RRF} \cite{DBLP:conf/sigir/CormackCB09}]
          [Voting-based
            [\textbf{BordaFuse} \cite{DBLP:conf/sigir/AslamM01}]
          ]
          [Probabilistic
            [\textbf{BayesFuse} \cite{DBLP:conf/sigir/AslamM01}]
          ]
        ]
        [Score-based
          [\textbf{WMNZ} \cite{DBLP:conf/cikm/WuC02}]
        ]
      ]
    \end{forest}
    }
    \caption{Overview of the analyzed data fusion methods.}
    \label{fig:data_fusion}
\end{figure}
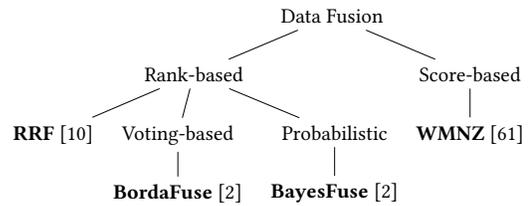

Data fusion is based on combining multiple rankings for a better overall retrieval performance than any single ranking out of the combined retrieval results would achieve \cite{DBLP:conf/jcdl/Efron09}. Generally, data fusion techniques can be categorized into \textit{rank-} and \textit{score-based} fusion methods \cite{DBLP:journals/jasis/LarsenIL09}, whereas particular \textit{rank-based} methods are described as \textit{voting-based} or \textit{probabilistic}~\cite{DBLP:conf/cikm/BassaniR22}. 
Figure \ref{fig:data_fusion} provides an overview of how data fusion approaches can be categorized and complements each of the four categories by the corresponding representative we use in our experimental setup.

Rank-based fusion methods combine multiple rankings based on the documents' rank positions. In this regard, these methods are score-independent, which is useful if score normalization is not an option, the score distributions of the single retrieval outputs are incompatible, or the scores are unavailable. For the most part of the data fusion experiments, we rely on Reciprocal Rank Fusion (RRF) \cite{DBLP:conf/sigir/CormackCB09} that is an entirely rank-based fusion method, which proved to be effective and robust. It is defined as follows:

\begin{equation}
  R R F \operatorname{score}(d \in D)=\sum_{r \in R} \frac{1}{k+r(d)}
  \label{eq:rrf}
\end{equation}

where $r(d)$ denotes the rank of a document $d$ in the set of documents $D$ out of the set of considered rankings $R$. The constant $k$ is set to $60$ per default \cite{DBLP:conf/sigir/CormackCB09}. Its simplicity lies in its independency of ranking scores, voting algorithms, or probabilistic methods compared to other data fusion approaches.


As a particular type of rank-based methods, voting-based fusion techniques apply voting algorithms for combining ranked lists of documents. For example, BordaFuse \cite{DBLP:conf/sigir/AslamM01} exploits the Borda Count algorithm, whereas the underlying principle treats retrieval systems as \textit{voters} and documents as \textit{candidates}. Each document is assigned a preference score (the vote), descending along the ranking. The combination of multiple rankings (and their voting preferences) can be seen as an analogy to multi-candidate election strategies.

As an alternative, rank-based methods can also be \textit{probabilistic}. For example, BayesFuse \cite{DBLP:conf/sigir/AslamM01} estimates the probability distribution of the relevance for each of the combined rankings and combines them by Bayesian inference. Based on prior knowledge about the relevance distribution, it requires sample rankings as a reference, which are used in a preceding training phase. As described below, we estimate these corresponding parameters from those rankings of the other years to avoid any data leakage in the final rankings. 

Contrary to rank-based methods, score-based fusion techniques require document rankings and corresponding retrieval scores. These fusion techniques combine single relevance scores to a final score in the fused ranking. Wu and Crestani~\cite{DBLP:conf/cikm/WuC02} introduced WMNZ as a weighted variant of CombMNZ \cite{DBLP:conf/trec/ShawF94}, which, in turn, is based on the idea of \textit{\textbf{m}ultiplying individual scores by the number of \textbf{n}on-\textbf{z}ero scores}. Compared to CombMNZ, WMNZ determines the score in the fused ranking by the sum of individual scores multiplied by the sum of weights for documents with non-zero scores.  

Even though many different data fusion methods exist, we see our selected methods as appropriate candidates for further analysis since they cover all four categories and are well-established in the community. That is also indicated by the citation count of the corresponding publications with Aslam and Montague (BordaFuse and BayesFuse) \cite{DBLP:conf/sigir/AslamM01} having over $420$ citations, Cormack et al. (RRF) \cite{DBLP:conf/sigir/CormackCB09} over $170$ citations, and Wu and Crestani (WMNZ) \cite{DBLP:conf/cikm/WuC02} over $40$ citations in the ACM Digital Library \cite{DBLP:conf/jcdl/GuhaSAL13} as of April 2023.

\section{Dataset}
\label{sec:dataset}

In this work we reuse the dataset from the broader medical domain that was created by Breuer et al.~\cite{DBLP:journals/scientometrics/BreuerST22}. It consists of the TREC Precision Medicine benchmarking dataset~\cite{DBLP:conf/trec/RobertsDVHBLP17,DBLP:conf/trec/RobertsDVHBL18,DBLP:conf/trec/RobertsDVHBLPM19} enriched with bibliographic information. 

\subsection{TREC Precision Medicine}
The Precision Medicine Track has been held at TREC in 2017~\cite{DBLP:conf/trec/RobertsDVHBLP17}, 2018~\cite{DBLP:conf/trec/RobertsDVHBL18} and 2019~\cite{DBLP:conf/trec/RobertsDVHBLPM19}\footnote{\url{http://www.trec-cds.org}, the track was also held in 2020 but that year is out of scope of the current analyses.}. It focuses on retrieval of literature on evidence-based treatments and clinical trials in the medical oncology domain in two tracks: \textit{1)} retrieval of scientific abstracts of papers containing treatments for patients, and \textit{2)} retrieval of clinical trials for patients. In 2019, a sub-task for the first task also included the specification up to three treatments recommended for a patient~\cite{DBLP:conf/trec/RobertsDVHBLPM19}. 
The goal of this track is to develop methods to retrieve fitting documents for patients' highly individual characteristics such as genetic mutations of their form of cancer in order for the patients receiving optimal treatment and care~\cite{nguyen_experimentation_2019,DBLP:conf/trec/RobertsDVHBLP17}.
Data source for the first task were PubMed/MEDLINE articles, for the second task the organisers of the task provided information on clinical trials\footnote{\url{ClinicalTrials.gov}}. For 2017 and 2018 the data has been identical~\cite{DBLP:conf/trec/RobertsDVHBLP17,DBLP:conf/trec/RobertsDVHBL18}, for 2019 both datasets have been extended~\cite{DBLP:conf/trec/RobertsDVHBLPM19}.
As the topics for the tasks, patient profiles including their disease, variant and demographic have been given. 
As an exception in 2017 the topics could also include other additional data.

We focus on task \textit{1)} without the sub-task. In each year, the submitting teams could submit up to 5 runs per task. Each run consists of a ranking of at most 1,000 paper IDs per topic. First, the pooled highest ranked papers were manually assessed by physician graduate students and postdocs at the National Library of Medicine in multiple dimensions with specific scales. Then, the categories were automatically merged into a singe three-level relevance score by which the retrieval effectiveness was determined~\cite{DBLP:conf/trec/RobertsDVHBLP17,DBLP:conf/trec/RobertsDVHBL18,DBLP:conf/trec/RobertsDVHBLPM19}. Table~\ref{tab:pm.submission_data} holds the number of relevance judgements, teams and runs for the scientific abstract task for the three years.

\begin{table}[t]
    \caption{Number of relevance judgements (qrels), of teams who submitted (teams) and of submitted runs (runs) per year.}

    \centering
    \resizebox{0.48\textwidth}{!}{
    \begin{tabular}{lll|lll|lll}
        \toprule

        \multicolumn{3}{c}{2017~\cite{DBLP:conf/trec/RobertsDVHBLP17}} & \multicolumn{3}{c}{2018~\cite{DBLP:conf/trec/RobertsDVHBL18}} & \multicolumn{3}{c}{2019~\cite{DBLP:conf/trec/RobertsDVHBLPM19}}\\
        \midrule

        Qrels & Teams & Runs & Qrels & Teams & Runs & Qrels & Teams & Runs\\
        \midrule

        22,642 &29 & 125 & 22,429 & 24 & 103 & 22,429 & 14 & 62\\
        \bottomrule

    \end{tabular}
    }
    \label{tab:pm.submission_data}
\end{table}

\subsection{Method Overview}

\begin{table}[t]
    \caption{Different retrieval engines, number of analyzed reports, approaches from the \ac{TREC-PM} 2017 to 2019.}

    \centering
    \begin{tabular}{llccc|c}

\toprule
	&& 2017	& 2018	& 2019	& $\sum$ \\
\midrule
&Reports per year	& 20	& 20	& 14	& 54 \\
\midrule
\parbox[t]{2mm}{\multirow{8}{*}{\rotatebox[origin=c]{90}{Engine}}} & ElasticSearch	& 5 &	8 &	7	& 20 \\
&Lucene	& 6	& 3	& 2	& 11 \\
&Terrier	& 3	& 3	& 1	& 7 \\
&unknown	& 1	& 2	& 2	& 5 \\
&Solr	& 2	& 2	& 1	& 5 \\
&Galago	& 2	& 	& 	& 2 \\
&Indri	& 1	& 1	& 	& 2 \\
&Whoosh 	& 	& 1	& 1	& 2 \\
\midrule
\parbox[t]{2mm}{\multirow{8}{*}{\rotatebox[origin=c]{90}{Approaches}}} & Query expansion	& 16	& 14	& 12	& 42 \\
&KB + ontologies	& 17	& 14	& 6	& 37 \\
&Re-ranking	& 6 & 7	& 9	& 22 \\
&Embeddings	& 3	& 5	& 5	& 13 \\
&Data fusion	& 4	& 5	& 3	& 12 \\
&LTR	& 1	& 3	& 5	& 9 \\
&LLM	&	&	& 3 &3 \\
&Citation-based 	& 2	& 	& 	& 2 \\
\bottomrule
    \end{tabular}
    \label{tab:trec-pm-statistics}
\end{table}

Nguyen et al. \cite{nguyen_experimentation_2019} provided an interactive tool to analyze different implementation setups and retrieval approaches for \ac{TREC-PM} 2017 and 2018. Later, Fässler et al. \cite{10.1145/3397271.3401048} analyzed the features that make a \ac{TREC-PM} engine successful. Using the optimization tool SMAC~\cite{SMAC} they analyzed over 100 retrieval parameters and found that the optimal combination can reach an infNDCG of 0.5732 and 0.6071 on previously unseen data for the biomedical abstracts and clinical trials tasks, respectively. These performance values are comparable to the best-performing systems in the three \ac{TREC-PM} editions 2017-2019. None of these tools and surveys focused on citation-based retrieval techniques and did not include them in their analyses. To complement this, we did a literature survey for all \ac{TREC-PM} publications for 2017-2019. 
Table~\ref{tab:trec-pm-statistics} gives an overview of the used retrieval engines and incorporated retrieval approaches of teams which submitted a report for \ac{TREC-PM} for our three considered years.
There is only information on \textit{if} a team used a retrieval engine and some retrieval approaches throughout any of their runs.
The retrieval engines do not differ between runs of the same team, the incorporated retrieval approaches, however, might.

\subsection{Specific Runs}
Only two teams used \textit{citation information} as part of their methods in a total of three runs. All of them were submitted in 2017: Team
CSIROmed~\cite{CSIRO} submitted the runs \texttt{aCSIROmedMGB} and \texttt{aCSIROmed\hyp{}PCB}. The first variant combines demographic attribute expansion with MeSH similarity re-ranking. The second run extends the first one by also expanding genes' and diseases' descriptions in the given topics. Both runs boost results based on citations received by clinical trials matching the given topic.
Team BiTeM~\cite{bitem} submitted the single run \texttt{SIBTMlit5}. For ranking abstracts to queries they combine citations from clinical trials with the type of article, e.g., it being a clinical trial or appearing in the proceedings of journals.

Some teams submitted plain BM25 runs throughout the years:
UKNLP~\cite{DBLP:conf/trec/NohK17} submitted run \texttt{UKY\_BASE} in 2017 which uses Lucene.
For 2018 team KlickLabs~\cite{DBLP:conf/trec/NishaniKB18} submitted run \texttt{KLPM18T2Bl}, using ElasticSearch. In the same year UCAS~\cite{DBLP:conf/trec/ZhengLHX18} used Terrier for their \texttt{UCASSA2} run.
In 2019 there were two runs; \texttt{BM25} incorporating ElasticSearch by the IMS Unipd~\cite{DBLP:conf/trec/NunzioMA19} team and \texttt{bm25\_6801} using Solr by CSIRO~\cite{DBLP:conf/trec/RybinskiKP19}.

\subsection{Merged Dataset}
For this paper, we reuse a dataset from the broader medical domain that was created by Breuer et al. \cite{DBLP:journals/scientometrics/BreuerST22}. Scientifically interesting about this dataset is that it contains relevance scores (non-relevant / fair / high) for scientific papers as well as corresponding bibliometric metadata. The dataset covers PubMed articles, for which relevance judgments from the \ac{TREC-PM} Abstract tasks and the corresponding documents were combined with their citation count in Web of Science and their Altmetric score from Altmetric.com. To obtain this dataset, the data from the different databases was matched using the PubMed ID, which is contained in all three sources. The dataset was created in 2020, covering 116,437 individual publications from 1942 to 2019. It is unique, because it contains different types of relevance (intellectual and cumulative relevance) for the same publications. The authors made the data publicly available on Zenodo~\cite{breuer_timo_2022_5883400}. Table \ref{tab:dataset.overview} reports the coverage of the bibliometric metadata for all three years. As can be seen, altmetrics has the lowest coverage. For most of the other bibliometric indicators, the metadata coverage is above 50\% with regards to the total number of judged documents.


\begin{table}[t]
\caption{
  Total number of unique documents for citations (C), altmetrics (A), publication year (P), research level (R) and impact factor (I). The percentage reports the relative amount regarding the total number of judged documents.
  }
\resizebox{0.49\textwidth}{!}{

\begin{tabular}{l|l|l|l|l|l}
\toprule
\textbf{Year} & \textbf{C} & \textbf{A} & \textbf{P} & \textbf{R} & \textbf{I}\\
\midrule
2017 & 14170 (66\%) & 6134 (29\%) & 14586 (68\%) & 14067 (66\%) & 11449 (53\%) \\
2018 & 11214 (55\%) & 4547 (22\%) & 11618 (57\%) & 11239 (55\%) & 9246 (45\%) \\
2019 & 11381 (61\%) & 5639 (30\%) & 12221 (66\%) & 11707 (63\%) & 9387 (51\%) \\
\bottomrule

\end{tabular}
}
\label{tab:dataset.overview}
\end{table}

\section{Methodology}
\label{sec:methodology}

\begin{figure}[!t]
    \centering
    \includegraphics[width=0.85\columnwidth]{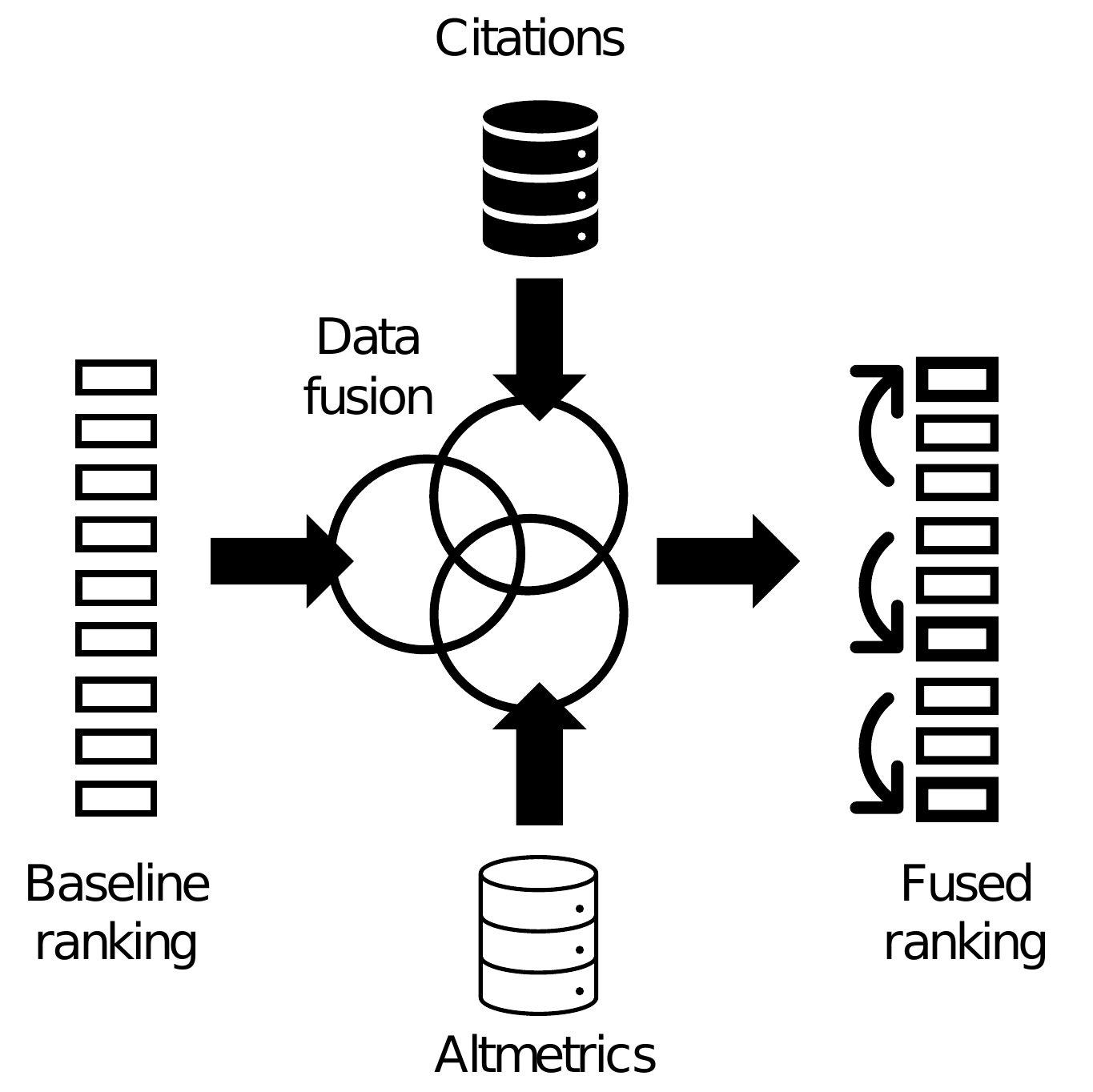}
    \caption{Methodology based on bibliometric data fusion of rankings and the principle of polyrepresentation.}
    \label{fig:methodology}
\end{figure}

The following experiments that give answers to our research questions (cf. Section \ref{sec:intro}) combine the run submissions of the \ac{TREC-PM} Abstract task from 2017 to 2019 with the bibliometric metadata based on the data fusion techniques introduced in the previous Section \ref{sec:data_fusion}. Figure \ref{fig:methodology} provides an overview of the general re-ranking approach based on data fusion. The original ranking is combined with bibliometric indicators such as citations and altmetrics. The data fusion operationalizes the principle of polypresentation and follows earlier work \cite{DBLP:conf/jcdl/LarsenI06,DBLP:journals/jasis/LarsenIL09}. The final ranking is a fused result based on the combination of multiple ranking criteria. As the retrieved result set still contains the results of the baseline ranking, the fused ranking can be considered a re-ranking, which can contribute to a better user experience \cite{DBLP:conf/jcdl/Nomoto12}.

Our first experiment evaluates the bibliometric relevance signals isolated from any of the \ac{TREC-PM} approaches. Besides exploiting the single bibliometric indicators as the ranking criteria, we also analyze all possible fused combinations. As a follow-up, we combine the \ac{TREC-PM} runs with the bibliometric metadata and evaluate the fused combinations as re-rankings compared to the original run versions, which were actually submitted to the shared task. Finally, we provide an outlook of how users could benefit from re-rankings based on bibliometric data fusion. For the most part, we rely on RRF, which proved to be a robust approach for meta-search and data fusion. However, to answer RQ2, we also include the three additional data fusion approaches to analyze the generalizability of our findings with other fusion algorithms.

The experiments are implemented with the help of the evaluation toolkit \texttt{ranx} \cite{DBLP:conf/ecir/Bassani22} and the corresponding support of data fusion methods \cite{DBLP:conf/cikm/BassaniR22}. For transparency and reproducibility, we provide open-source releases of the code and experimental data\footnote{\faGithub \ \url{https://github.com/irgroup/jcdl2023-data-fusion}}. The experiments can be reproduced in an interactive Jupyter Notebook that, for instance, can be rerun on Google Colab. The \ac{TREC-PM} runs must be retrieved from a password-protected section on the TREC website. We do not provide them as an additional data resource as we respect the intellectual property of the authors who submitted their experimental results to the shared tasks. However, we provide snapshots of the directory trees regarding the file location of the runs in order to repeat the experiments with as much ease and rigor as possible.   

\section{Experimental Results}
\label{sec:experimental_results}

In the following, we present the experimental results and give answers to our research questions.

\subsection{RQ1: Bibliometric Relevance Signals}
\label{sec:bibliometric.relevance.signals}

\begin{table}[t]
\caption{
  Retrieval effectiveness: the best bibliometric results are highlighted in boldface. Superscripts denote significant differences in Fisher's Randomization Test~\cite{fisher1936design} with $p \le 0.05$ when comparing bibliometric rankings.
  }
\resizebox{0.47\textwidth}{!}{

\begin{tabular}{l|l|l|l|l|l|l|l}
\toprule

 & \textbf{Model} & \textbf{C} & \textbf{A} & \textbf{P} & \textbf{R} & \textbf{I} & \textbf{BM25} \\

\midrule

\parbox[t]{2mm}{\multirow{5}{*}{\rotatebox[origin=c]{90}{\textbf{2017}}}} & Recall & 0.7853$^{ARI}$ & 0.4162 & \textbf{0.7972}$^{CARI}$ & 0.7608$^{AI}$ & 0.6301$^{A}$ & 0.4640 \\
& nDCG & 0.4992$^{ARI}$ & 0.3163 & \textbf{0.5069}$^{ARI}$ & 0.4666$^{AI}$ & 0.4162$^{A}$ & 0.4423 \\
& AP & \textbf{0.1812}$^{AI}$ & 0.1020 & 0.1733$^{AI}$ & 0.1546$^{A}$ & 0.1399$^{A}$ & 0.1636 \\
& P$@$10 & \textbf{0.2700}$^{R}$ & 0.2400$^{R}$ & 0.2033 & 0.1200 & 0.2500$^{R}$ & 0.4667 \\
& Bpref & \textbf{0.1577} & 0.1434 & 0.1541 & 0.1307 & 0.1444 & 0.2714 \\

\midrule

\parbox[t]{2mm}{\multirow{5}{*}{\rotatebox[origin=c]{90}{\textbf{2018}}}} & Recall & 0.7916$^{ARI}$ & 0.4066 & \textbf{0.8019}$^{CARI}$ & 0.7739$^{AI}$ & 0.6438$^{A}$ & 0.7828 \\
& nDCG & \textbf{0.5728}$^{ARI}$ & 0.3651 & 0.5671$^{ARI}$ & 0.5297$^{AI}$ & 0.4744$^{A}$ & 0.6376 \\
& AP & \textbf{0.2905}$^{ARI}$ & 0.1765 & 0.2815$^{AI}$ & 0.2591$^{AI}$ & 0.2261$^{A}$ & 0.3195 \\
& P$@$10 & 0.3760$^{R}$ & \textbf{0.3860}$^{R}$ & 0.3180$^{R}$ & 0.2360 & 0.3420$^{R}$ & 0.5680 \\
& Bpref & \textbf{0.2896}$^{AI}$ & 0.2355 & 0.2809$^{A}$ & 0.2612 & 0.2506 & 0.4852 \\

\midrule

\parbox[t]{2mm}{\multirow{5}{*}{\rotatebox[origin=c]{90}{\textbf{2019}}}} & Recall & 0.8260$^{AI}$ & 0.4732 & \textbf{0.8849}$^{CARI}$ & 0.8435$^{AI}$ & 0.6690$^{A}$ & 0.7574 \\
& nDCG & 0.5754$^{ARI}$ & 0.3693 & \textbf{0.6031}$^{ARI}$ & 0.5433$^{AI}$ & 0.4818$^{A}$ & 0.5870 \\
& AP & 0.2756$^{ARI}$ & 0.1633 & \textbf{0.2896}$^{ARI}$ & 0.2442$^{A}$ & 0.2182$^{A}$ & 0.2584 \\
& P$@$10 & \textbf{0.3525}$^{RI}$ & 0.2850$^{R}$ & 0.3075$^{R}$ & 0.1925 & 0.2850$^{R}$ & 0.5125 \\
& Bpref & \textbf{0.2460}$^{R}$ & 0.2064 & 0.2416 & 0.2024 & 0.2283 & 0.3946 \\

\bottomrule

\end{tabular}
}
\label{tab:bibliometric.relevance.signals}
\end{table}

In the following, we investigate RQ1: \emph{To what extent can bibliometric relevance signals be used as ranking criteria for biomedical information retrieval?} This evaluation captures single signals and combinations of signals' effectiveness in standard information retrieval measures.

\subsubsection{Setting}
To this end, we use the bibliometric metadata described earlier as follows. For all five types of bibliometric metadata, including citations (C), altmetrics (A), publication year (P), research level (R), and impact factor (I), we rank the documents by the corresponding count in decreasing order. Consequently, documents with higher citations and altmetrics, research levels, and impact factors are 
more relevant.

Regarding the publication year, we rank more recent publications higher than older ones, i.e., the ranking implies a recency-based criterion. Similarly, the publications are ranked by a decreasing research level, where higher levels correspond to more basic research and lower levels to more applied research contributions. Finally, note that all rankings are query-agnostic, i.e., the documents are ranked by the five bibliometric indicators, independent of the topic or a corresponding query. 

For better comparison, we include selected BM25 runs mentioned earlier, i.e., we select one default BM25 implementation for each year. We include \texttt{UKY\_BASE} \cite{DBLP:conf/trec/NohK17} for 2017, \texttt{UCASSA2} \cite{DBLP:conf/trec/ZhengLHX18} for 2018, and \texttt{BM25} \cite{DBLP:conf/trec/NunzioMA19} for 2019. Two runs are based on Lucene's implementations of BM25 \cite{DBLP:conf/trec/NohK17,DBLP:conf/trec/NunzioMA19}, while the third \cite{DBLP:conf/trec/ZhengLHX18} from 2018 is based on the Terrier retrieval toolkit. We acknowledge that there may be differences between BM25 implementations \cite{DBLP:conf/ecir/KamphuisVBL20}. However, these runs guarantee the best comparability, as they use default settings and do not use query expansion techniques.

\subsubsection{Single Signals}
Table \ref{tab:bibliometric.relevance.signals} compares the five different types of bibliometric metadata when used for ranking medical abstracts of the \ac{TREC-PM} tasks in terms of the recall rate, \ac{nDCG} \cite{DBLP:journals/tois/JarvelinK02}, \ac{AP} \cite{DBLP:books/daglib/0021593}, \ac{P@10} \cite{DBLP:books/daglib/0021593}, and Bpref \cite{DBLP:conf/sigir/BuckleyV04} scores. Unless stated otherwise, all measures are evaluated with a cut-off value of 1,000. As shown in Table \ref{tab:dataset.overview}, the metadata information about the publication year has the highest coverage. This circumstance leads to higher recall rates of the publication year for all three years. Generally, the metadata with a higher coverage also results in higher nDCG and AP scores as they are recall-dependent. Generally, the coverage of altmetrics is low, as can also be seen by the low recall rates and the overall lowest nDCG and AP scores. However, the rankings based on altmetrics achieve comparably good results for P@10. For instance, in 2018, they achieve the best results, while ranking third after citations and the publication year for the 2017 and 2019 tasks. As highlighted by the bold numbers, the rankings based on citations, altmetrics, and the publication year achieve the best results in all three years.

When comparing the bibliometrics rankings to the BM25 runs, we see that some bibliometric indicators can achieve higher recall rates, which also explains the higher nDCG and AP scores. However, the BM25 runs outperform the bibliometric indicators in terms of P@10 and Bpref.

\begin{figure}[]
  \centering
  \includegraphics[width=0.8675\columnwidth]{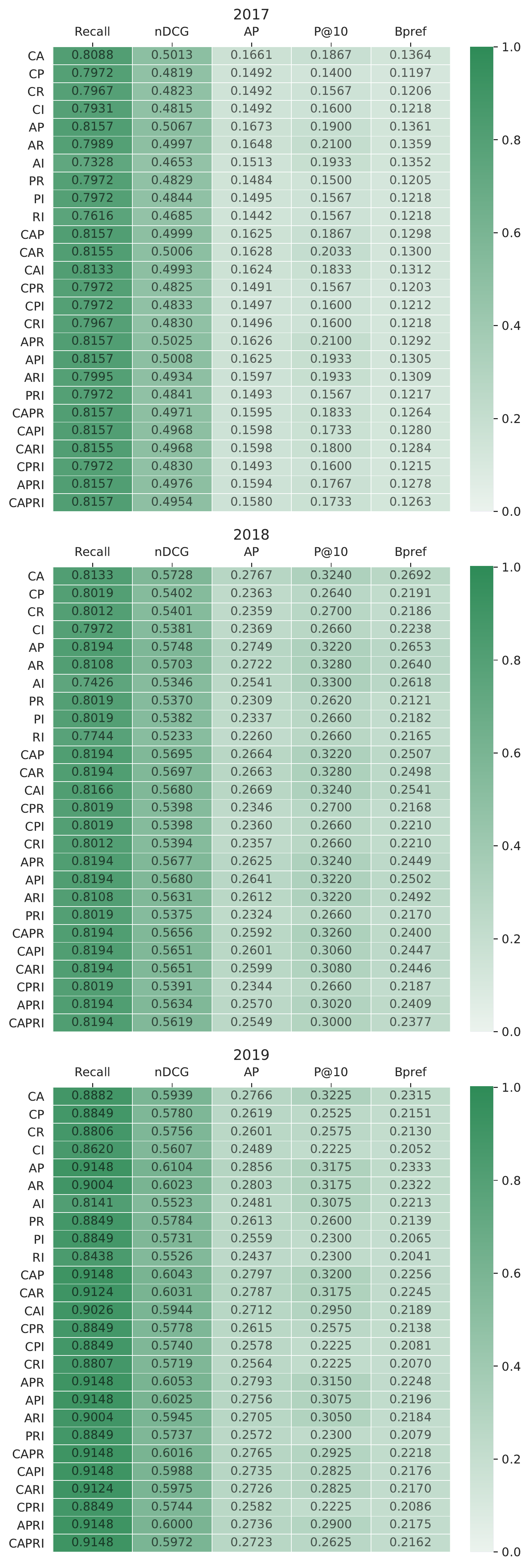}
  \caption{Retrieval effectiveness of fused bibliometric signals including all possible combinations for \ac{TREC-PM} 2017-2019.}
  \label{fig:capri.rrf.heatmap}
\end{figure}

\subsubsection{Fused Signals}
Before combining the \ac{TREC-PM} runs with the bibliometric metadata, we evaluate fused bibliometric-based rankings in isolation. To this end, we determine all possible combinations for the five different types of bibliometric data and evaluate them by the recall rates, nDCG, AP, P@10, and Bpref. Figure \ref{fig:capri.rrf.heatmap} shows the results for all possible RRF-based combinations evaluated by the \ac{TREC-PM} relevance judgments of all three years. As can be seen from the heatmaps, the combinations yield different performance scores and score variability. 

On the one hand, we confirm that data fusion can lead to improved retrieval results. However, some fused rankings yield higher recall rates than single ranking criteria in Table \ref{tab:bibliometric.relevance.signals}. For instance, the highest recall rate of 0.8157 (see Fig.~\ref{fig:capri.rrf.heatmap}) is achieved by multiple fused combinations (including AP, CAP, APR, API, CAPR, CAPI, APRI, CAPRI). It is slightly higher than the best recall rate of 0.7972 based on the publication year for \ac{TREC-PM} 2017 (see Tab.~\ref{tab:bibliometric.relevance.signals}). Similarly, the recall can be improved by some fused combinations and is higher than that of single ranking criteria for the other two years. 

On the other hand, fused combinations can also harm retrieval performance, as can be seen by the other measures. For instance, none of the fused results can outperform the best P@10 scores by single ranking criteria reported in Table \ref{tab:bibliometric.relevance.signals}. For \ac{TREC-PM} 2017, the best-fused combinations (AR, APR) result in P@10=0.2100, while the best result is based on citations (C) with P@10=0.2700. Similarly, most fused combinations stay below the best results by the single ranking criteria in Table \ref{tab:bibliometric.relevance.signals} regarding nDCG, AP, and Bpref. 

\subsubsection{Discussion}
\textbf{In conclusion, citations, altmetrics, and the publication year deliver the best results when used as separate ranking criteria as shown in Table \ref{tab:bibliometric.relevance.signals}}. Our fused combinations of the different metadata types have shown that recall rates primarily improve to a moderate extent. However, most of the fused combinations did not yield a better retrieval performance than that of the individual metadata types. Once again, the rankings are independent of any query or topic-related information. We think a reasonable ranking should consider topic- or query-related information as part of the data fusion. Suppose the fused combinations are only based on the query-agnostic bibliometric metadata. In that case, the rankings possibly drift away from the topical relatedness. 

Nonetheless, as an answer to RQ1, we see promise in the \textit{signal strength} of bibliometric metadata. When used as separate ranking criteria, the metadata types yield acceptable retrieval results that possibly provide additional relevance signals, which could complement other ranking approaches. 
Overall, these first outcomes motivate us to combine the metadata with other ranking methods that were used in \ac{TREC-PM}. In the following, we use citations, altmetrics, and the publication year for data fusion with the runs of \ac{TREC-PM} 2017-2019. 

\subsection{RQ2: Overall Retrieval Performance}

\begin{figure*}[!t]
  \centering
  \includegraphics[width=.925\textwidth]{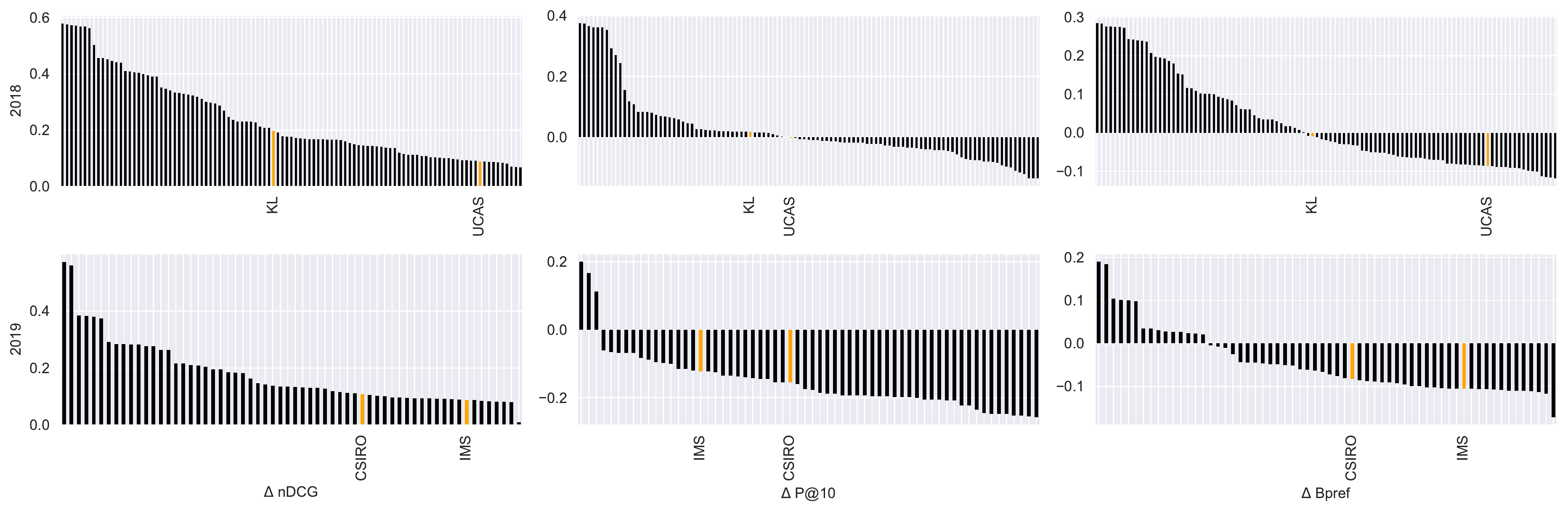}
  \caption{Rank fusion-based improvements over the baseline runs for the TREC Precision Medicine Abstract task for 2018 and 2019. BM25 runs marked in orange and named according to an abbreviation of the team's name.}
  \label{fig:delta.rrf}
\end{figure*}

\begin{figure*}[!t]
  \centering
  \includegraphics[width=.925\textwidth]{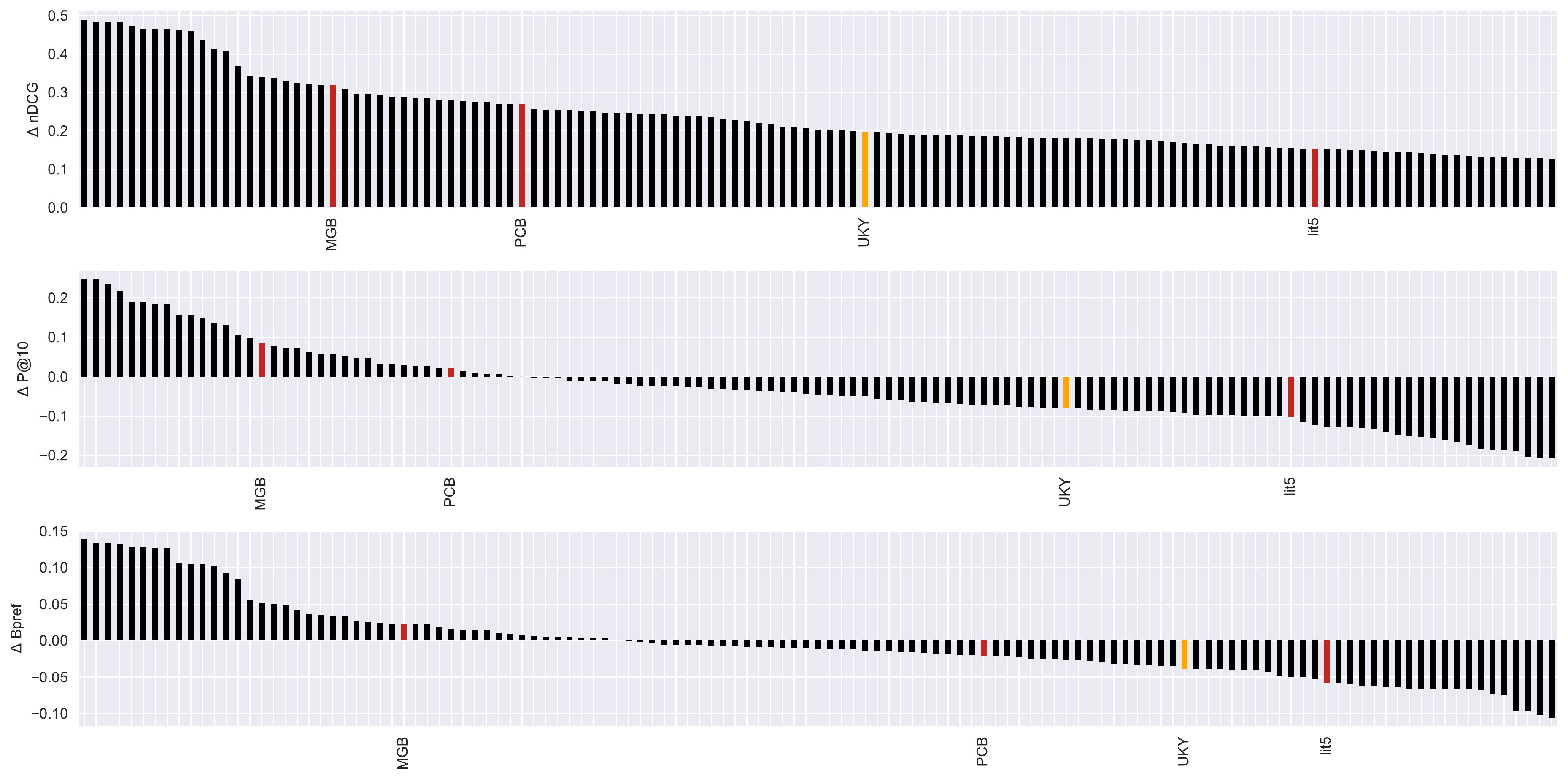}
  \caption{Rank fusion-based improvements over the baseline runs for the \ac{TREC-PM} Abstract task for 2017 with methods using citations (MGB: aCSIROmedMGB, PCB: aCSIROmedPCB, lit5: SIBTMlit5) marked in red. BM25 run marked in orange.}
  \label{fig:delta17.rrf}
\end{figure*}

\begin{table*}[!ht]
    \centering
    \caption{
      Comparison of different rank fusion methods for \ac{TREC-PM} from 2017 to 2019.
      } 
    \resizebox{\textwidth}{!}{
    \begin{tabular}{lrrr|rrr|rrr|rrr}
      \toprule
      {} & \multicolumn{3}{c}{Reciprocal Rank Fusion (RRF) \cite{DBLP:conf/sigir/CormackCB09}} & \multicolumn{3}{c}{BordaFuse \cite{DBLP:conf/sigir/AslamM01}} &  \multicolumn{3}{c}{BayesFuse \cite{DBLP:conf/sigir/AslamM01}} & \multicolumn{3}{c}{WMNZ \cite{DBLP:conf/cikm/WuC02}} \\ 
\midrule
{} &        2017 &        2018 &      2019 &        2017 &        2018 &      2019 &        2017 &        2018 &      2019 &        2017 &        2018 &      2019 \\
\midrule
Number of systems               &         125 &         103 &        62 &         125 &         103 &        62 &         125 &         103 &        62 &         125 &         103 &        62 \\
\midrule
(Signif.*) improvements (nDCG)  &  125 / 125* &  103 / 103* &  62 / 61* &  125 / 125* &  103 / 103* &  62 / 61* &  125 / 125* &  103 / 103* &  62 / 62* &  125 / 125* &  103 / 103* &  62 / 62* \\
Average improvement (nDCG)      &      0.2378 &      0.2384 &    0.1815 &      0.2248 &      0.2286 &    0.2001 &      0.2198 &      0.2334 &    0.2205 &      0.2366 &      0.2418 &     0.243 \\
Overall change (nDCG)           &      0.2378 &      0.2384 &    0.1787 &      0.2248 &      0.2286 &    0.1975 &      0.2198 &      0.2334 &    0.2205 &      0.2366 &      0.2418 &     0.243 \\
\midrule
(Signif.*) improvements (AP)    &  125 / 123* &  103 / 103* &  62 / 55* &  125 / 114* &  103 / 103* &  62 / 62* &  125 / 120* &  103 / 103* &  62 / 62* &  123 / 121* &  103 / 103* &  62 / 62* \\
Average improvement (AP)        &      0.1173 &      0.1849 &    0.1237 &      0.1132 &      0.1727 &    0.1482 &      0.1065 &      0.1815 &    0.1772 &      0.1265 &      0.2025 &    0.2036 \\
Overall change (AP)             &      0.1163 &      0.1849 &    0.1161 &      0.1073 &      0.1727 &    0.1482 &      0.1041 &      0.1815 &    0.1772 &      0.1225 &      0.2025 &    0.2036 \\
\midrule
(Signif.*) improvements (P@10)  &    37 / 18* &    46 / 19* &    3 / 3* &    41 / 14* &    39 / 18* &   10 / 1* &     31 / 9* &    31 / 19* &   11 / 1* &    44 / 16* &    60 / 24* &   35 / 4* \\
Average improvement (P@10)      &      0.1589 &      0.2221 &      0.16 &      0.1112 &      0.1747 &     0.295 &      0.0696 &      0.1887 &      0.28 &      0.1104 &      0.1636 &    0.1312 \\
Overall change (P@10)           &     -0.0299 &      0.0223 &   -0.1518 &     -0.0393 &      -0.003 &   -0.0728 &     -0.0463 &      0.0004 &   -0.0373 &     -0.0267 &      0.0294 &     0.015 \\
\midrule
(Signif.*) improvements (Bpref) &    46 / 17* &    47 / 36* &   15 / 6* &    39 / 13* &    44 / 36* &  21 / 11* &    34 / 14* &    46 / 37* &  27 / 17* &    74 / 25* &    61 / 45* &  61 / 44* \\
Average improvement (Bpref)     &      0.1047 &      0.1668 &    0.1294 &      0.0968 &       0.134 &    0.0925 &      0.0642 &      0.1446 &    0.1023 &      0.0834 &      0.1481 &    0.0786 \\
Overall change (Bpref)          &     -0.0033 &      0.0244 &   -0.0453 &     -0.0141 &      0.0123 &   -0.0124 &      -0.017 &      0.0196 &    0.0162 &      0.0106 &      0.0553 &    0.0635 \\
\bottomrule
      \end{tabular}
      }
      \label{tab:retrieval.performance}
  \end{table*}

\begin{figure*}[!tb]
  \centering
  \includegraphics[width=0.9\textwidth]{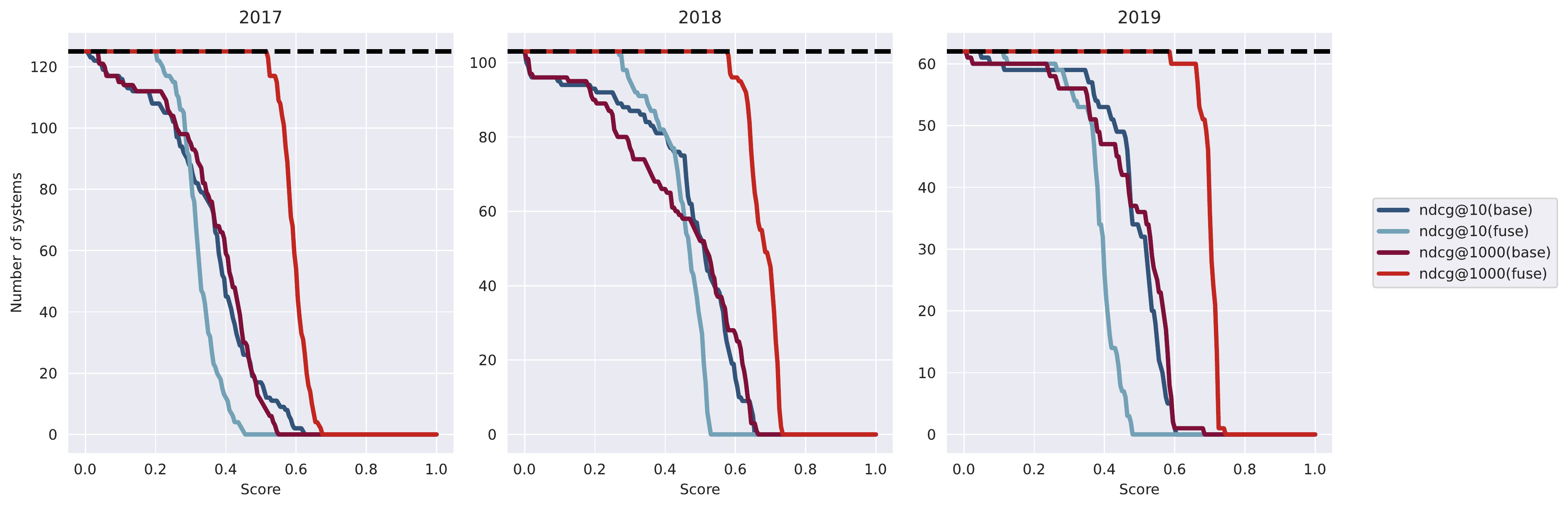}
  \caption{Number of systems vs. retrieval effectiveness before (dark) and after (light) bibliometric data fusion for nDCG@10 (blue) and nDCG@1000 (red) for \ac{TREC-PM} from 2017 to 2019. The dashed line corresponds to the total number of systems.}
  \label{fig:ndcg.rrf}
\end{figure*}

Here we address RQ2: \emph{Can bibliometric-enhanced data fusion methods improve the overall retrieval performance?} 

\subsubsection{Difference due to Data Fusion}
Figure \ref{fig:delta.rrf} shows the differences between the run submissions before and after the bibliometric data fusion for the run submission of the \ac{TREC-PM} Abstract task from 2018 and 2019. The bar plots show the differences between each run, i.e., the baseline and the re-ranked results after the bibliometric data fusion based on RRF using all features (CAPRI\footnote{The order of features is irrelevant, they are all applied at the same time.}). Figure \ref{fig:delta17.rrf} shows the rank fusion-based improvements for 2017 and marks methods using citations as part of their approach in red.

\textbf{For all systems, the nDCG (and AP) scores can be improved.} Regarding the P@10 and Bpref scores, only a fraction of the systems improve, and for the majority of systems, the retrieval performance decreases. In order to put these results into a larger context and to draw conclusions about the generalizability, Table \ref{tab:retrieval.performance} compares different rank fusion methods for all three tracks from 2017 to 2019. In addition to RRF, the Table includes the other data fusion techniques introduced in Section \ref{sec:data_fusion}. In contrast to the other three approaches, BayesFuse requires a training phase to parameterize the probability distribution of the relevance, for which we make use of the run submissions from the other years to avoid target data leakage in the evaluations. For instance, we use a sample of five runs from 2018 to optimize the parameters for the data fusion with the runs from 2017. Similarly, we use sample runs from 2017 and 2019 to optimize the data fusion for the other years.

When considering the three citation-based runs from 2017 (see the red marks in Figure~\ref{fig:delta17.rrf}) we can only find improvements for \texttt{aCSIROmedMGB} in all metrics. The other two runs' performance does not consistently improve over all three measures.
Both citation-based runs by CSIROmed use the same core components. The difference is \texttt{aCSIROmedPCB}'s additional incorporation of expansions of the topic's fields. Although seemingly unintuitive, the smaller improvement of the more complex method could be attributed to it, achieving a higher performance independent of data fusion~\cite{CSIRO}. 

When again considering the five BM25 runs for comparability through the three years (see the orange marks in Figures~\ref{fig:delta.rrf} and \ref{fig:delta17.rrf}), we see some differences between them. We encounter improvements in recall as shown by the nDCG values, but worse precision as P@10 and Bpref mainly decrease when using rank fusion.

In general, the earlier results are confirmed on a larger scale. While there are some differences between the data fusion methods, the nDCG and AP scores generally improve for most runs, whereas there are deteriorated P@10 and Bpref scores. Furthermore, Table \ref{tab:retrieval.performance} includes the number of significant differences between the baselines and the fused run versions. The results show that most improvements are significant for nDCG and AP. In addition, Table \ref{tab:retrieval.performance} includes the average improvement based on the significant\footnote{Fisher's Randomization Test, $p \le 0.05$} differences as well as the overall change that is determined with all differences between the baselines and the fused runs. Overall, we see that the improvements in the nDCG and AP scores can be generalized over the three different tracks from 2017 to 2019, with nearly all improvements being significant. 

\subsubsection{Change in nDCG}
It is fair to criticize that most improvements could be attributed to weak baselines with low nDCG scores. For this reason, Figure \ref{fig:ndcg.rrf} shows a more detailed analysis of the nDCG scores. All plots show the total number of systems above the nDCG scores (with cut-offs at position 10 or 1,000) on the x-axis for the three years of \ac{TREC-PM}. The leftmost plot shows the distributions of systems submitted to TREC Precision Medicine 2017. The dashed horizontal line corresponds to the total number of run submissions (125). As can be seen, the best-performing systems achieved an nDCG@10 score above 0.6 (cf. to the dark blue line plot in Figure \ref{fig:ndcg.rrf}), while all systems had an nDCG@1000 score lower than 0.6 (cf. to the dark red line plot in Figure \ref{fig:ndcg.rrf}). 

In comparison, the lighter-colored line plots show the distribution of systems after the bibliometric data fusion. As can be seen, all of the systems have an nDCG@1000 score above 0.5 after data fusion (cf. to the light red line plot in Figure \ref{fig:ndcg.rrf}). Likewise, the best nDCG@1000 scores are above 0.6. In contrast, the best nDCG@10 score is below 0.5 after data fusion (cf. to the light blue line plot in Figure \ref{fig:ndcg.rrf}), which complies with the earlier findings. As can be seen from the precision-oriented measures, there are fewer relevant results in the top-ranked positions, and the Precision and nDCG@10 scores deteriorate from the bibliometric data fusion. For the other two years, the same outcomes can be observed.

\begin{figure}[!t]
  \centering
  \includegraphics[width=.8\columnwidth]{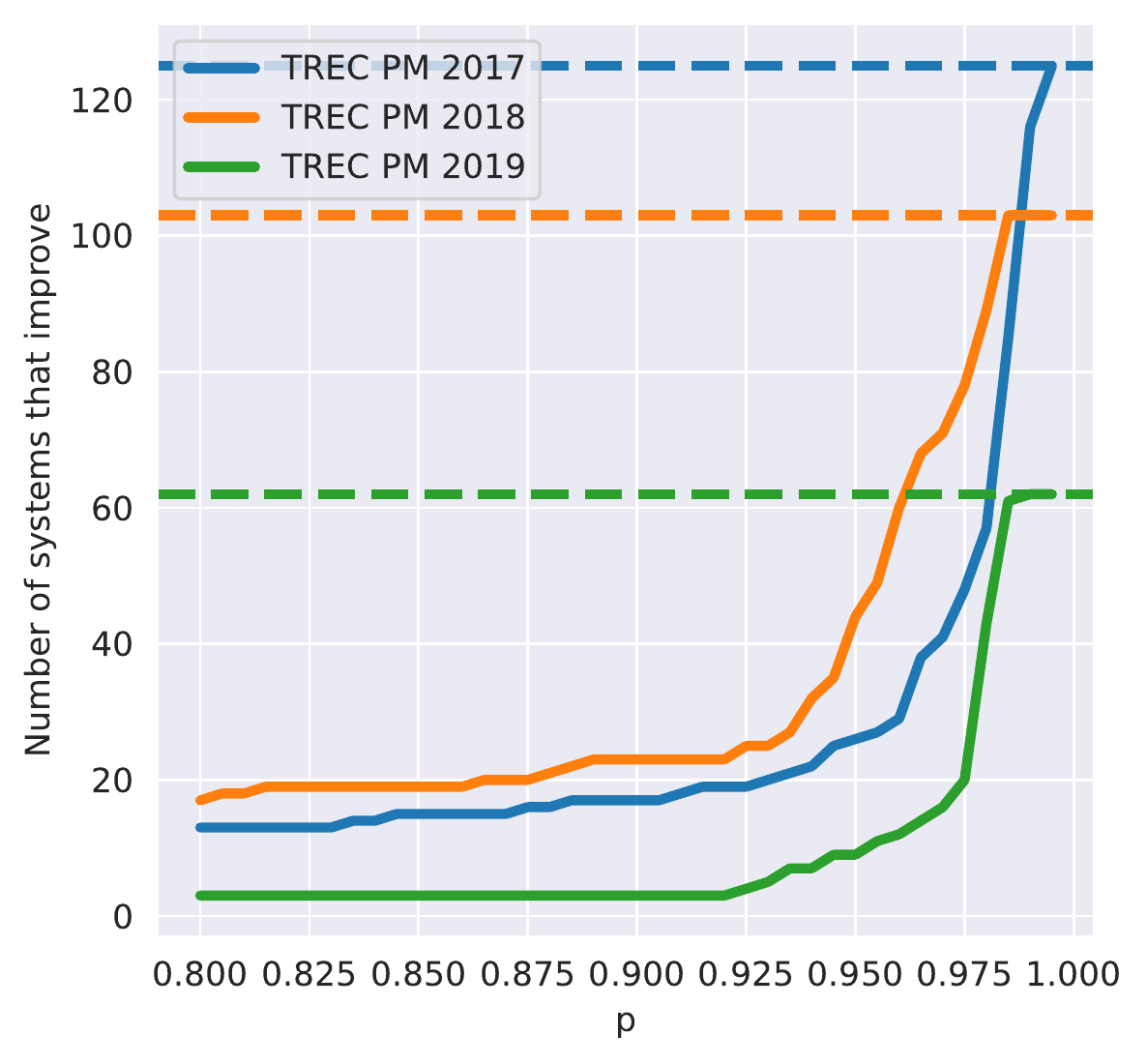}
  \caption{Number of system improvements vs. user persistence for \ac{TREC-PM}. $p$ models the transition probability to the next document of RBP.}
  \label{fig:rbp.rrf}
\end{figure}

\subsubsection{Implications for Users}

Finally, we address how users benefit from bibliometric-enhanced data fusion results. We tackle this research question by investigating the Rank-biased Precision (RBP) as a realistic approximation of users' patience \cite{DBLP:journals/tois/MoffatZ08}, which is defined as follows:

\begin{equation}
  \label{eq:rbp}
  \mathrm{RBP}=(1-p) \cdot \sum_{i=1}^d r_i \cdot p^{i-1}
\end{equation}

where $d$ denotes the total number of documents in a ranking and $r_i$ the document's relevance at rank $i$. The parameter $p$ models the transition probability to the next ranked document and models the user's patience: the larger $p$, the higher the probability of inspecting the next document in the ranking and the more patient the user. Compared to nDCG and AP, RBP does not require knowledge about the recall, which is a more realistic assumption about the user, according to Moffat and Zobel \cite{DBLP:journals/tois/MoffatZ08}.


For all three tracks from 2017 to 2019, we compare the number of improved systems over $p$. The corresponding results are shown in Figure \ref{fig:rbp.rrf}. Again, the dashed lines correspond to the total number of runs that were submitted in each particular year, and that could potentially be improved. We see that the overall improvement, i.e., the number of improved systems, is generally lower for more impatient users ($p=0.8$), whereas the benefit is higher as the user gets more patient. Especially for $p > 0.975$, there is a steep increase in the number of systems that improve for all three tracks. 

\subsubsection{Discussion}
\textbf{In conclusion, this analysis reconfirms the following two results. First, not only weak baselines but also better-performing systems benefit from the data fusion, as is also underlined by the significance tests in Table \ref{tab:retrieval.performance}. Second, the improvements are mainly useful for recall-oriented tasks, as already suggested by the outcomes of the previous experiments (cf. Section \ref{sec:bibliometric.relevance.signals}).} The nDCG scores at higher cut-offs mainly improve, while nDCG scores with lower cut-offs, e.g., the nDCG@10 scores, are generally lower after data fusion. Our RQ2, which considers the overall improvement of the retrieval performance, is answered positively. The performance improves (significantly) when using bibliometric-enhanced rank fusion, especially the recall. The RBP-based evaluations showed that \textbf{the more patient the user, the higher the benefit of bibliometric-enhanced data fusion approaches}.

\subsection{Discussion}

This work is an example of the successful implementation of science models into academic retrieval processes discussed by Mutschke et al. \cite{DBLP:journals/scientometrics/MutschkeMSS11}. who promoted the idea of using retrieval experiments as a kind of litmus test to evaluate the plausibility of models on scientific communication and scientific collaboration. Our earlier work showed that bibliometrics correlate to a certain extent with editorial relevance judgments of IR test collections~\cite{DBLP:journals/scientometrics/BreuerST22}. Furthermore, bibliometrics can be considered implicit relevance indicators representing relevance from diverse perspectives. Therefore, data fusion is a feasible solution to combine multiple relevance signals~\cite{DBLP:journals/jasis/LarsenIL09}, and the underlying principles align with the concept of polyrepresentation~\cite{ingwersen_cognitive_1996}. To this end, we analyzed how bibliometric data fusion techniques can be used to rank biomedical abstracts.

Our experiments showed that bibliometric indicators could be exploited for ranking medical abstracts to a moderate extent when used as \textit{single signals}, i.e., without any topic-related ranking criteria. Furthermore, they achieved reasonable recall rates that depend on the overall coverage of the bibliometric metadata regarding the judged documents. These results comply with earlier work that showed the overlap between editorial relevance labels and bibliometrics~\cite{DBLP:journals/scientometrics/BreuerST22}. Conversely, the precision rates showed that single signals did not retrieve many relevant abstracts at the top ranks. These trends became even clearer when the single bibliometric indicators were combined in a preliminary data fusion experiment. The RRF-based \textit{fused signals}, including all possible bibliometric metadata combinations, even deteriorated the precision rates. Most of the retrieval outcomes were worse than that of single bibliometric signals. While the recall rates slightly improved, we conclude from these experiments that the fused bibliometric signals further \textit{drift away} from reasonable rankings. 

We conclude that bibliometric indicators imply a shallow or more implicit notion of relevance that can only reveal its potential when combined with topic-related ranking criteria, as it was demonstrated by the data fusion experiments with the \ac{TREC-PM} runs.
Bibliometric data fusion has a primarily recall-enhancing effect. For all systems, nDCG and AP could be improved when comparing the fused rankings to the original run submissions. Regarding precision, most systems deteriorate, and only for a small fraction of the submitted runs the data fusion improves the performance.
 
There were slightly better results based on the evaluations with Bpref - a measure that excludes unjudged documents. These outcomes indicate that bibliometric data fusion also \textit{brings up} documents that were not part of the pooling to compile candidates for the relevance judgments. As a future perspective, we emphasize the importance of including bibliometric metadata in the submissions as part of shared tasks. Our review of the \ac{TREC-PM} tasks from 2017 to 2019 revealed that only a minority of participants (two groups) used bibliometric indicators in the rankings. It is possible to improve the evaluation setup by harnessing such relevance indicators. It makes the pooled set of abstracts more diverse and allows a better evaluation of bibliometric-based rankings.

Finally, we analyzed how users of digital libraries would benefit from bibliometric data fusion. The corresponding experiments simulated users with different levels of patience. We implemented these experiments with the help of the RBP measure, which precisely considers this aspect. While an impatient user,  who browses through the top-ranked abstracts, will not recognize much improvement in the retrieval results, a more patient user will likely benefit from bibliometric data fusion. As our experiments showed, improved fused systems increased as the simulated user gets more patient. We highlight that curators of academic digital libraries should consider these findings. If it is possible to classify the search behavior of users, for instance, by their interaction patterns \cite{DBLP:conf/jcdl/LiuS22}, re-ranking the results with bibliometric data fusion approaches can improve the user experience.

\section{Conclusion}
\label{sec:conclusion}

In summary, data fusion in information retrieval can effectively identify relevant documents by combining the output of multiple models. The strength of this approach is that it can push relevant documents ranked lower on individual output lists to higher positions in the fused results. While finding relevant abstracts using only bibliometric indicators is generally possible, there are low precision rates as topic-related ranking criteria are not considered. We propose bibliometric data fusion with runs from TREC Precision Medicine as a solution. Our evaluations showed that bibliometric relevance signals could improve retrieval performance. 


As part of future work, it would be interesting to analyze bibliometric data fusion in other scientific domains. Scientific disciplines differ in their scholarly communication habits, which impacts citations or altmetrics. Thus, it is required to investigate how well our findings generalize with other data. In general, we see bibliometric indicators as valuable relevance-bearing information that should be further investigated. For instance, it is possible to distinguish between important and less important citations, which could be used to weight single citations differently. Similarly, citation networks of authors that go beyond the raw count of citations could be considered. Finally, our results suggest that users of digital libraries could benefit from bibliometric data fusion. The outcomes of our simulated experiments should be validated in user studies that analyze the impact in real-world environments. 

\balance
\bibliographystyle{ACM-Reference-Format}

\end{document}

%% file: acronyms.tex
\acrodef{TREC}[TREC]{Text REtrieval Conference}
\acrodef{TREC-PM}[TREC-PM]{TREC Precision Medicine}
\acrodef{nDCG}[nDCG]{normalized Discounted Cumulated Gain}
\acrodef{AP}[AP]{Average Precision}
\acrodef{P@10}[P@10]{Precision at 10}